\documentclass[conference]{IEEEtran}
\IEEEoverridecommandlockouts
\usepackage{cite}
\usepackage{amsmath,amssymb,amsfonts}
\usepackage{algorithmic}
\usepackage{graphicx}
\usepackage{textcomp}
\usepackage{xcolor}
\usepackage{acronym}
\acrodef{MPC}{Model Predictive Control}
\acrodef{EMPC}{Economic Model Predictive Control}
\acrodef{OCP}{Optimal Control Problem}
\acrodef{NLP}{Nonlinear Program}
\acrodef{SDP}{Semidefinite Program}
\acrodef{MDP}{Markov Decision Process}
\acrodef{KKT}{Karush-Kuhn-Tucker}
\acrodef{IFT}{Implicit Function Theorem}
\acrodef{RL}{Reinforcement Learning}
\acrodef{DRL}{Deep Reinforcement Learning}
\acrodef{MLP}{Multi-Layer Perceptron}
\acrodef{NN}{Neural Network}
\acrodef{ML}{Machine Learning}
\acrodef{DNN}{Deep Neural Network}
\acrodef{DPG}{Deterministic Policy Gradient}
\acrodef{TD3}{Twin Delayed Deep Deterministic Policy Gradient}
\acrodef{DQN}{Deep Q-Network}
\acrodef{SQP}{Sequential Quadratic Programming}
\acrodef{QP}{Quadratic Programming}
\acrodef{iLQR}{iterative Linear Quadratic Regulator}

\usepackage{tikz}
\usepackage{pgfplots}
\usepackage{pgfplotstable} 
\usepackage{booktabs} 
\usepackage{adjustbox}

\usepackage{svg}

\usepackage{hyperref}

\usepackage{afterpage}
\usepackage{geometry}
\usepgfplotslibrary{groupplots}
\usepgfplotslibrary{statistics} 
\pgfplotsset{compat=1.18}

\usetikzlibrary{arrows, positioning, calc, fit}
\def\BibTeX{{\rm B\kern-.05em{\sc i\kern-.025em b}\kern-.08em
T\kern-.1667em\lower.7ex\hbox{E}\kern-.125emX}}

\newcommand{\secref}[1]{Sec.~\ref{#1}}
\newcommand{\figref}[1]{Fig.~\ref{#1}}

\newcommand{\argmin}{\mathop{\mathrm{arg\, min}}\limits}

\begin{document}


\newcommand{\acados}{\texttt{acados}}
\newcommand{\numpy}{\texttt{numpy}}
\newcommand{\scipy}{\texttt{scipy}}
\newcommand{\CasADi}{\texttt{CasADi}}
\newcommand{\CleanRL}{\texttt{CleanRL}}
\newcommand{\Gymnasium}{\texttt{Gymnasium}}
\newcommand{\Python}{\texttt{Python}}
\newcommand{\qpOASES}{\texttt{qpOASES}}
\newcommand{\Ipopt}{\texttt{IPOPT}}
\newcommand{\HPIPM}{\texttt{HPIPM}}
\newcommand{\rlmpc}{\texttt{MPC4RL}}
\newcommand{\leapc}{\texttt{leap-c}}

\newcommand{\diffmpc}{\texttt{differentiable-mpc}}
\newcommand{\PyTorch}{\texttt{PyTorch}}
\newcommand{\stablebaselines}{\texttt{stable-baselines3}}

\title{\rlmpc\, -  A Software Package for Reinforcement Learning based on Model Predictive Control\\
   \thanks{
      $^1$ Department of Engineering Cybernetics, Norwegian University of Science and Technology, 7034 Trondheim, Norway\\
      $^2$ Department of Microsystems Engineering, University Freiburg, 79110 Freiburg, Germany\\
      This research was supported by NFR via SARLEM (project 300172), and by DFG via Research Unit FOR 2401, project 424107692, 504452366 (SPP 2364), and 525018088, by BMWK via 03EI4057A and 03EN3054B, and by the EU via ELO-X 953348
      (Corresponding author's e-mail: dirk.p.reinhardt@ntnu.no).
   }
}

\author{\IEEEauthorblockN{Dirk Reinhardt\textsuperscript{1}, Katrin Baumgärtner\textsuperscript{2}, Jonathan Frey\textsuperscript{2}, Moritz Diehl\textsuperscript{2}, Sebastien Gros\textsuperscript{1}}
}

\maketitle

\begin{abstract}
   In this paper, we present an early software integrating Reinforcement Learning (RL) with Model Predictive Control
   (MPC). Our aim is to make recent theoretical contributions from the literature more accessible to both the RL and MPC
   communities. We combine standard software tools developed by the RL community, such as \Gymnasium, \stablebaselines,
   or \CleanRL\, with the \acados\, toolbox, a widely-used software package for efficient MPC algorithms. Our core
   contribution is \rlmpc, an open-source \Python\, package that supports learning-enhanced MPC schemes for existing
   \acados\, implementations. The package is designed to be modular, extensible, and user-friendly, facilitating the
   tuning of MPC algorithms for a broad range of control problems. It is available on GitHub.
\end{abstract}

\vspace{0.5cm}



\section{Introduction}

\subsection{Motivation and Background}

\ac{MPC} has emerged as a pivotal technology in both industrial and academic spheres, known for its ability to handle
complex control problems under constraints. \ac{EMPC}~\cite{faulwasser2018economic} further enhances this by incorporating economic performance
directly into the control objective, making it highly relevant in any system where economic objectives naturally occur,
e.g. resource allocation problems, energy-saving technologies or process industries. In parallel, \ac{RL} has seen a
surge in interest due to its potential in learning optimal control policies through interaction with the environment.
The fusion of \ac{MPC} and \ac{RL} (among other \ac{ML} tools) is a rapidly evolving topic, promising enhanced
performance in varying, stochastic, and complex environments~\cite{gros2020datadriven,zanon2021safe}.

Despite the theoretical advancements and the growing body of academic literature in this area, there remains a
noticeable gap in the provision of open-source software tools for \ac{RL} based on \ac{MPC}. We aim at closing this gap
with a software package \rlmpc\, which is part of \leapc. Open-source software is necessary for reproducibility and transparency in
research. Furthermore, the implementation of the methods enabling \rlmpc\, is not trivial and the lack of software tools
to support implementations is a significant barrier to the widespread adoption of \ac{MPC} and \ac{RL} in academia (and
industry). Filling that gap should accelerate developments in this area, for which we present the first version of an
open-source \Python\, package~\footnote{\url{https://github.com/leap-c/leap-c}}.

The package is designed as a tool for tuning of \ac{MPC} algorithms using \ac{RL} techniques. From the perspective of
the \ac{ML} community, \ac{MPC} can be seen as a structured function approximator. The RL part of the package
is built around popular libraries such as \Gymnasium~\cite{towers2023gymnasium} and
\texttt{stable-baselines3}~\cite{raffin2021stablebaselines3} or \CleanRL~\cite{huang2022cleanrl} for building environments and interfacing with \ac{RL}
algorithms. The MPC part of the package is built around the software tools \CasADi~\cite{andersson2019casadi}
and \acados~\cite{verschueren2022acados}, which help to build \ac{MPC} problems easily and enable the use of
efficient \ac{OCP} solvers to implement them in practice. We aim at a design that is close to the \ac{RL} libraries,
such that it is easy to use and extend.

Similar projects focusing on implementations that establish a connection between \ac{RL} and \ac{MPC} exist. A
contribution from the machine-learning community is \diffmpc~\cite{amos2018differentiable}, which provides an \ac{MPC}
module that can be integrated into a larger \ac{RL} pipeline, as e.g. in the context of
robotics~\cite{romero2024actorcritic}.  However, it is limited to \ac{iLQR} as the underlying \ac{OCP} solver and can
only handle box constraints on the inputs.  In the control community, a similar project is available on
GitHub\footnote{\url{https://github.com/FilippoAiraldi/mpc-reinforcement-learning/}}. It is wrapping general-purpose
\ac{NLP} solver interfaces to build and evaluate \ac{MPC} schemes and uses custom implementations of \ac{RL} algorithms.
The main shortcoming of this approach is that it does not enable tight integration with the underlying \ac{OCP} solver,
which is crucial for the efficient computation of policy gradients.  In contrast, our approach to building a
framework that connects with state-of-the-art software, such as \acados, results in significantly improved computational
efficiency, as we demonstrate with numerical results in \secref{sec:case_studies}. We are also convinced that using
adopted libraries lowers the access barrier for the respective users, and facilitates development and testing by active
communities in both \ac{MPC} and \ac{RL}.

\newgeometry{
   top=0.75in,
   bottom=0.75in,
   left=0.75in,
   right=0.75in
}

\subsection{Outline and Contributions}

The core contribution of this paper is the \rlmpc\, package. It provides a set of customizable tools to compute
sensitivities of \ac{OCP} solutions required by \ac{RL} algorithms that use them as function approximators. We extend
\acados\, to provide the sensitivities along with the solution of the \ac{OCP}, which enables fast computation of the
corresponding policy gradient. Our results show that the evaluation of the policy gradient is about one magnitude faster
than general-purpose libraries that are commonly used to compute the sensitivities outside the \ac{OCP}. Considering the
importance of policy gradients in \ac{RL} and other learning domains that also enter problems within the control
community, we believe this software to be a significant contribution. The package is available on GitHub, and we aim to
expand this tool over time.

The outline of the paper is as follows. In \secref{sec:background}, we provide background on core concepts of the
connection between \ac{RL} and \ac{MPC}. In \secref{sec:module}, we discuss the design of the \rlmpc\, package and provide
an example in \secref{sec:case_studies} to demonstrate the use of the package. We outline future work in
\secref{sec:future_work} and give some concluding remarks in~\secref{sec:conclusion}.

\section{Background on Reinforcement Learning based on Model Predictive Control}
\label{sec:background}

In this section, we provide a brief overview of the core concepts of the connection between \ac{RL} and \ac{MPC}. The
equations outlined in this section allow for custom implementations of \ac{RL} algorithms based on \ac{MPC} with the
required value functions, policy, and their respective sensitivities. The \rlmpc\, package aims at integrating these
concepts into a user-friendly and efficient software package with \acados\, providing the \ac{OCP} solutions and the respective sensitivities in a fast and efficient way.

\subsection{Markov Decision Processes}

The \ac{MDP} provides a generic framework for the class of problems at the center of \ac{MPC}. An \ac{MDP} operates over
given state and action (aka input) spaces $S, A$, respectively. These spaces can be both discrete (i.e. integer),
continuous, or mixed sets. An \ac{MDP} is defined by the triplet $(L,\gamma, \rho)$, where $L$ is a stage cost,
$\gamma\in(0,1]$ a discount factor and $\rho$ a conditional probability (measure) defining the dynamics of the system
considered, i.e. for a given state-action pair $ s, a\in S\times A$, the successive state $s_+$ is distributed according
to
\begin{align}
   \label{eq:dyn_mdp}
   s_+ \sim \rho(\cdot | s, a)
\end{align}
Note that this is equivalent to the classic dynamics, deterministic or not, often considered in MPC, usually cast as
\begin{align}
   \label{eq:dyn_classic}
   s_+ =  F\left( s, a, w \right),\quad  w\sim W,
\end{align}
where $w$ is a random disturbance from the distribution $W$. In the special case $ w=0$, this simply yields deterministic dynamics. An
MDP is the problem of finding an optimal policy $ \pi^\star\,:\, S\rightarrow A$ as a solution of:
\begin{equation}
   \pi^\star \in \argmin_{\pi}\, J( \pi),
\end{equation}
with
\begin{equation}
   J( \pi)   = \mathbb{E}\left[\left. \sum_{k=0}^\infty\, \gamma^k L\left( s_k, a_k\right)\,\right|\,  a_k= \pi\left( s_k\right)\right],
\end{equation}
and the expected value operator $\mathbb{E}[.]$ is taken over the (possibly) stochastic closed-loop trajectories of the system.
Discussing the solution of \acp{MDP} is often best done via the Bellman equations defining implicitly the optimal value function
$V^\star\,:\, S\rightarrow \mathbb R$ and the optimal action-value function $Q^\star\,:\, S\times A\rightarrow \mathbb R$ as
\begin{subequations}
   \label{eq:Bellman0}
   \begin{align}
      V^\star\left( s\right)    & =\min_{ a}\, Q^\star\left( s, a\right) \label{eq:Bellman0:V}                               \\
      Q^\star\left( s, a\right) & = L\left( s, a\right) + \gamma \mathbb E\left[V^\star\left( s_+\right)\,|\, s, a\, \right]
   \end{align}
\end{subequations}
An optimal policy then reads:
\begin{align}
   \label{eq:OptPolicy0}
   \pi^\star\left( s\right) \in\argmin_{a}\, Q^\star\left( s, a\right)
\end{align}

\subsection{Reinforcement Learning}\label{sec:RL:Intro}
The fundamental goal of Reinforcement Learning (RL) is to use data to deliver an approximation of an optimal policy $
   \pi^\star$. The field can be coarsely divided into two large classes of approaches. The first class, which we refer to
as value-based, approximates the optimal action-value function $Q^\star$ via a parameterized function approximator
$Q_{\theta}$. The parameters $\theta$ are then adjusted using data such that $Q_{\theta}\approx Q^\star$ in some sense.
An approximation of an optimal policy $ \pi^\star$ can then be obtained using:
\begin{align}
   \label{eq:OptPolicy:Approx}
   \hat{ \pi}^\star\left( s\right) =\argmin_{ a}\, Q_{\theta}\left( s, a\right)
\end{align}
The second class approximates $ \pi^\star$ directly via a parametrized policy $ \pi_{\theta}$, and adjusts the
parameters $\theta$ from data so as to minimize $J( \pi_{\theta})$. This can, e.g., be done by estimating policy
gradients $\nabla_{\theta}J( \pi_{\theta})$, or by building surrogate models of $J( \pi_{\theta})$, used to adjust
$\theta$. This class can be referred to as policy-based methods.

\subsection{Model Predictive Control}

The theoretical developments in~\cite{gros2020datadriven} are the basis for the \rlmpc\, package. We briefly summarize the main results here.
An \ac{EMPC} problem that is parameterized by the parameter vector $\theta$ can be formulated as
\begin{subequations}
   \label{eq:V_NLP}
   \begin{align}
      V_\theta (s) = \min _{x,u,\sigma }\ \  & V^\mathrm{f}_\theta (x_N) + \rho^\mathrm{f}(\sigma _N)
      + \sum _{k=0}^{N-1} \ell_\theta (x_k,u_k) + \rho(\sigma_k) \label{eq:V_cost}                                     \\
      \text{s.t.} \ \                        & x_0 = s\label{eq:V_constraint0}                                         \\
                                             & x_{k+1} = f_\theta \left(x_k,u_k\right)\label{eq:V_constraint1}         \\
                                             & g\left(u_k\right) \leq 0 \label{eq:V_constraint2}                       \\
                                             & h_\theta \left(x_k,u_k,\sigma _k\right) \leq 0 \label{eq:V_constraint3} \\
                                             & h^\mathrm{f}_\theta (x_N,\sigma _N) \leq 0 \label{eq:V_constraint4}
   \end{align}
\end{subequations}
Note that we distinguish between $s,a$ as the state-action pair that corresponds to the current state and the action that is
taken, and $x,u$ as the state-action pair that corresponds to the state and action that form the predictions in the \ac{NLP}.

The problem with prediction horizon $N$ includes the stage cost $\ell_\theta$ and terminal cost $V^\mathrm{f}_\theta$ in~\eqref{eq:V_cost}. The
initial state $x_0 = s$ in~\eqref{eq:V_constraint0} is propagated through the dynamic model $f_\theta$ in~\eqref{eq:V_constraint1} with path inequality constraint
$h_\theta$ in~\eqref{eq:V_constraint3} and the terminal constraint $h^\mathrm{f}_\theta$ in~\eqref{eq:V_constraint4}, all parameterized through
$\theta$. The path inequality constraints at the intermediate stages and at the terminal stage may be relaxed via
penalty functions $\rho,\,\rho^\mathrm{f}$ acting on the respective slack variables $\sigma_k,\,\sigma_N$. The pure input constraints $g$
in~\eqref{eq:V_constraint2} are not relaxed or parameterized, considering that they are typically simple polytopic
constraints that are known in practice. Recent results in~\cite{kordabad2024equivalence} established the equivalence of undiscounted
\ac{MPC} and discounted \ac{MDP} problems and alleviate the need for discounting in the \ac{EMPC} formulation. However,
it is straightforward to include discounting in the \ac{EMPC} formulation and the following discussion~\cite{gros2020datadriven}.

In standard \ac{MPC} schemes, the \ac{NLP}~\eqref{eq:V_NLP} is solved at each time step. The first element of the optimal input
sequence $u^\star = (u_0^\star,\ldots,u_{N-1}^\star)$ is applied to the system, i.e. the policy is given by
\begin{equation}
   \pi _\theta (s) = u_0^\star
\end{equation}
The solution of the \ac{NLP}~\eqref{eq:V_NLP} is thus a function of the current state $s$ and the parameter $\theta$,
and delivers the policy $\pi _\theta (s)$ and the value function $V_\theta (s)$. To approximate the state-action value
function $Q_\theta (s,a)$, we consider a modified \ac{NLP} that is obtained by fixing the first input
$u_0=a$:
\begin{subequations}
   \label{eq:Q_NLP}
   \begin{align}
      Q_\theta (s,a) = \min _{u,x,\sigma} & \quad \hbox{\eqref{eq:V_cost}}                                            \\
      \text{s.t.}                         & \quad \hbox{\eqref{eq:V_constraint1}}\!-\!\hbox{\eqref{eq:V_constraint3}} \\
                                          & \quad u_0 = a
   \end{align}
\end{subequations}
The definitions of the policy $\pi _\theta (s)$, the state value function $V_\theta (s)$, and the state-action value function
satisfy the Bellman identities
\begin{equation}
   \pi _\theta (s) = \argmin _a\, Q_\theta (s,a),\quad V_\theta (s) = \min _a\, Q_\theta (s,a)
\end{equation}
Note that $f_\theta$ is a deterministic parametric model of the true system dynamics $F$, and $\ell_\theta$ is a
parametric stage cost which is not necessarily aligned with the true stage cost $L$. The combination \ac{RL} and \ac{MPC} is
then to learn the parameters $\theta$ such that the \ac{NLP} solutions approximate the optimal value function and an
optimal policy. The functions $\ell_\theta$ and $f_\theta$ should thus ideally be chosen such that the overall \ac{MPC}
scheme can model the \ac{MDP}. In practice, considerations of system identification, model complexity, and computational
efficiency will guide the choice of the parametric models.

\subsection{Sensitivities of NLP Solutions in Reinforcement Learning}
\ac{RL} algorithms naturally rely on the value function, action-value function, and policy underlying the Bellman
identities. State-of-the-art \ac{RL} algorithms moreover operate in gradient-based fashion to update these quantities,
which means that they require the respective sensitivities $\nabla_\theta V, \nabla_\theta Q, \nabla_\theta \pi$. In
standard \ac{NN} architectures, building the sensitivities is straightforward. When using \ac{OCP} solvers as function
approximations, this task is less trivial as it requires differentiating \ac{NLP} solutions. We discuss next how to obtain the
sensitivities in this case by differentiating the corresponding \ac{NLP} at the solution.

Using the primal variables of the NLPs~\eqref{eq:V_NLP},\eqref{eq:Q_NLP}, $x=(x_0,\ldots,x_N)$ and
$u=(u_0,\ldots,u_{N-1})$, and the slack variables $\sigma=(\sigma_0,\ldots,\sigma_N)$, the Lagrange function of
\eqref{eq:Q_NLP} reads
\begin{align}
   \mathcal {L}_\theta & (s,a,y) = \sum _{k=0}^{N-1} \ell_\theta (x_k,u_k)+ \rho(\sigma _k) +V^\mathrm{f}_\theta (x_N) + \rho^\mathrm{f}(\sigma _N) \nonumber                   \\
                       & + \chi _0^\top \left(x_0 - s\right) + \sum _{k=0}^{N-1} \chi _{k+1}^\top \left(f_\theta\left(x_k,u_k\right) - x_{k+1} \right)\nonumber                 \\
                       & + \zeta ^\top (u_0-a) + \sum _{k=0}^{N-1} \nu _k^\top g\left(u_k\right)  \nonumber                                                                     \\
                       & + \sum _{k=0}^{N-1} \mu_{k}^\top h_\theta \left(x_k,u_k,\sigma_k\right) + \mu _N^\top h^\mathrm{f}_\theta (x_N,\sigma_N) \label{eq:lagrange_function},
\end{align}
where the dual variable are given by the Lagrange multipliers $\chi=(\chi_0,\ldots,\chi_{{N-1}})$ associated with the equality
constraints~\eqref{eq:V_constraint1} and $\mu=(\mu_0,\ldots,\mu_{N-1})$ associated with the mixed inequality
constraints~\eqref{eq:V_constraint2}. The dual variables $\nu=(\nu_0,\ldots,\nu_{N-1})$ are associated with the pure input
constraints~\eqref{eq:V_constraint3}, and the dual variable $\zeta$ corresponds to the initial input constraint $u_0=a$
in~\eqref{eq:Q_NLP}. Note that at $u_0=a$ or $\zeta=0$, the Lagrange functions of \eqref{eq:V_NLP} and \eqref{eq:Q_NLP} coincide.
For a shorter notation, we collect the primal-dual variables in the vector $y=(x,u,\sigma,\chi,\mu,\nu,\zeta)$.
Moreover, at the primal-dual solution $y^\star$ of the \acp{NLP}, the gradients of the Lagrange function with respect to the parameters $\theta$
deliver the gradients of the state value function $V_\theta (s)$ and the state-action value function $Q_\theta (s,a)$:
\begin{subequations}
   \label{eq:V_Q_gradient}
   \begin{align}
      \nabla_\theta
      V_\theta (s)= \nabla_\theta \mathcal{L}_\theta
      (s,u^\star_0,y^\star), \\	\nabla_\theta Q_\theta (s,a)=
      \nabla_\theta \mathcal{L}_\theta (s,a,y^\star)
   \end{align}
\end{subequations}
The gradient of the policy $\pi _\theta (s)$ can be obtained by differentiating the \ac{KKT} conditions of the
\ac{NLP}~\eqref{eq:V_NLP} or \eqref{eq:Q_NLP} which are satisfied at their respective solutions $y^*$. Consider the common case where \ac{NLP} solutions
are delivered via \ac{SQP} with an interior-point method to solve the \ac{QP} approximations and let $\tau$ denote the
associated homotopy parameter. Then the system of equations solved within an interior-point \ac{QP} solver reads:
\begin{subequations}
   \label{eq:KKT}
   \begin{align}
      \nabla_{x,u,\sigma} \mathcal{L}_\theta =                  & \ 0 \label{eq:Stationarity}                    \\
      x_0 - s =                                                 & \ 0                                            \\
      u_0-a =                                                   & \ 0 \label{eq:initial_action_constraint}       \\
      f_\theta\left(x_k,u_k\right) - x_{k+1}  =                 & \ 0,                                           \\
      g(u_k) + t_{\nu_k, i} =                                   & \ 0,  \label{eq:ComplementarySlackness1}       \\
      h_\theta(x_k, u_k, \sigma_k) + t_{\mu_k, i} =             & \ 0,  \label{eq:ComplementarySlackness2}       \\
      h^\mathrm{f}_\theta (x_N,\sigma_N) + t_{\mu_N, i} =       & \ 0,                                           \\
      \nu_{k,i} t_{\nu_k, i}     - \tau                       = & \ 0, \quad \forall i \in \{1,\dots, n_{g_k}\}, \\
      \mu_{k,i} t_{\mu_k, i}     - \tau                      =  & \ 0, \quad \forall i \in \{1,\dots, n_{h_k}\}, \\
      \mu_{k,N} t_{\mu_N, i}     - \tau                      =  & \ 0, \quad \forall i \in \{1,\dots, n_{h_N}\},
   \end{align}
\end{subequations}
where $n_{g_k}, n_{h_k}, n_{h_N}$ are the number of constraints at stage $k$ and the terminal stage, respectively, and
$t_{\nu_k, i}, t_{\mu_k, i}, t_{\mu_N, i}$ are the slack variables corresponding to the inequalities~\cite{Frison2020a}.

Note that \eqref{eq:initial_action_constraint} and the corresponding term in \eqref{eq:lagrange_function} are only
included when the \ac{MPC} approximates the action-value function.  Collecting $\eqref{eq:KKT}$ as $\xi_\theta = 0$, the
policy gradient can then be obtained by using the \ac{IFT} for evaluating parametric sensitivities of \ac{NLP}
\begin{equation}
   \nabla _{\theta }\pi _\theta (s) = - \nabla_{\theta } \xi_\theta (s,y^\star) \nabla _{y} \xi _\theta (s,y^\star)^{-1} \frac{\partial y}{\partial u_0} \label{eq:policy_gradient}
\end{equation}

Utilizing the results summarized in this section allows to connect the \ac{MPC} scheme~\eqref{eq:V_NLP} to \ac{RL} algorithms
that train deterministic policies such as Q-Learning as found in \cite{sutton2018reinforcement}, \ac{DPG} by \cite{silver2014deterministic}
or \ac{TD3} by \cite{fujimoto2018addressing}.

\section{The \rlmpc\, Package}
\label{sec:module}

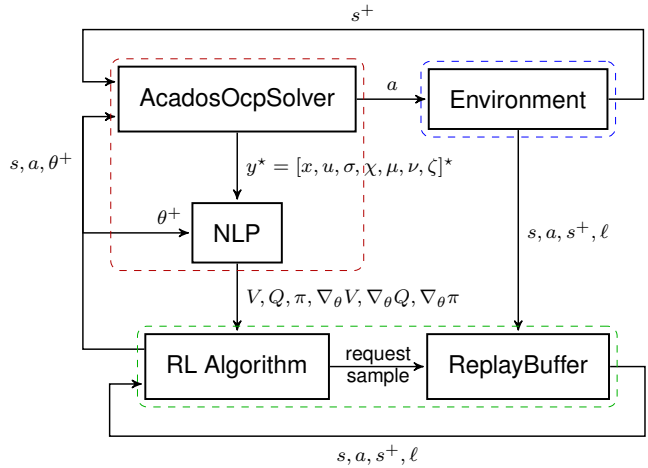
\begin{figure}[htbp]
   \centering
   \begin{adjustbox}{max width=\columnwidth}
      \begin{tikzpicture}[
        font=\sffamily,
        every matrix/.style={ampersand replacement=\&,column sep=1cm,row sep=1cm},
        source/.style={draw,thick,rounded corners,fill=yellow!20,inner sep=.3cm},
        process/.style={draw,thick,circle,fill=blue!20},
        sink/.style={source,fill=green!20},
        module/.style={draw,thick,rectangle,inner sep=.3cm},
        datastore/.style={draw,very thick,shape=datastore,inner sep=.3cm},
        dots/.style={gray,scale=2},
        to/.style={->,>=stealth',shorten >=1pt,semithick,font=\sffamily\footnotesize},
        every node/.style={align=center},
        fitbox/.style={draw, dashed, inner sep=0.1cm, rectangle, rounded corners}]

    \matrix{
        \node[module] (solver) {AcadosOcpSolver}; \& \node[module] (env) {Environment}; \\
        \node[module, opacity=0.4] (nlp) {NLP}; \& ;                                    \\
        \node[module] (rl) {RL Algorithm}; \& \node[module] (buffer) {ReplayBuffer};    \\
    };

    \coordinate[left=of nlp, shift={(-1cm, 0cm)}] (lofnlp) {};
    \coordinate[right=of env, shift={(-0.5cm, 1cm)}] (rofenv);
    \coordinate[left=of solver, shift={(+0.5cm, 1cm)}] (lofsolver);
    \coordinate[left=of solver, shift={(+0.5cm, -0.25cm)}] (lofsolver2);
    \coordinate[right=of buffer, shift={(-0.5cm, -1cm)}] (rofbuffer);
    \coordinate[left=of rl, shift={(+0.5cm, -1cm)}] (lofrl);

    \draw[to] (solver) -- node[midway,right] {$y^\star=[x,u,\sigma,\chi,\mu,\nu,\zeta]^\star$} (nlp);

    \draw[to] (nlp) -- node[midway,right] {$V, Q, \pi, \nabla_{\theta}V, \nabla_{\theta}Q, \nabla_{\theta}\pi$} (rl);

    \draw[to] ([shift={(0cm,+0.25cm)}]rl.west) -| node[pos=0.9, left] {$s, a, \theta^+$} (lofsolver2) -- ([shift={(0cm,-0.25cm)}]solver.west);

    \draw[to] (lofsolver2) |- node[pos=0.9, above] {$\theta^+$} (nlp.west);

    \draw[to] (env.east) -| (rofenv) -- node[midway, above] {$s^+$} (lofsolver) |- ([shift={(0cm,0.25cm)}]solver.west);

    \draw[to] (buffer.east) -| (rofbuffer) -- node[midway, below] {$s, a, s^+, \ell$} (lofrl) |-([shift={(0cm,-0.25cm)}]rl.west);

    \draw[to] (solver.east) -- node[midway, above] {$a$} (env.west);

    \draw[to] (env.south) -- node[midway,right] {$s, a, s^+, \ell$} (buffer.north);

    \draw[to] (rl.east) -- node[midway] {request\\sample} (buffer.west);

    \node[fitbox, color=black!30!red, fit=(solver) (nlp)] {};
    \node[fitbox, color=black!30!green, fit=(rl) (buffer)] {};
    \node[fitbox, color=blue, fit=(env)] {};

\end{tikzpicture}
   \end{adjustbox}
   \caption{Communication between the main components of the \rlmpc\, package. The \texttt{AcadosOcpSolver} communicates
      the primal-dual solution to the NLP class, which computes the sensitivities of the NLP solutions. The RL algorithm
      requests samples from the replay buffer and updates the parameters of the \texttt{AcadosOcpSolver} (policy). The
      environment communicates the state-action-cost transitions to the replay buffer. The different colors indicate the
      external code bases used to implement the different components with \CasADi/\acados\, (red),
      \stablebaselines\, (green), and \Gymnasium\, (blue). The NLP object replicating the NLP structure is used to compute
      the sensitivities of the NLP solutions. It is not needed when computing the sensitivities with \acados.}
   \label{fig:interface}
\end{figure}

In this section, we discuss the design of the \rlmpc\, package, its main components, and how they interact. We use established
libraries, and provide a high-level overview of the communication between the main components in \figref{fig:interface}.

\subsection{Model Predictive Control Implementations}

Communities in industry and academia have developed a wide range of tools to implement MPC schemes. Popular tools
include \CasADi~\cite{andersson2019casadi} to build nonlinear symbolic expression and generate compiled code for the
corresponding functions and their derivatives.  If runtime is not a concern, it is sufficient to use interfaces to
standard solver packages such as \Ipopt~\cite{wachter2006implementation}.  An overview on software packages for embedded
optimization, with respective application targets and licences is given in~\cite{verschueren2022acados}.  If runtime is
a concern, efficient \ac{OCP} solution algorithms can be obtained using structure exploiting algorithms and compiled
code, as can be done using \acados~\cite{verschueren2022acados}. Efficient solvers for MPC are essential for tackling
challenging applications where the real-time constraint is demanding. In the context of \rlmpc, efficient solvers are
also highly beneficial during the training phase regardless of the physical sampling time of the real system. Indeed, if
the training of the MPC involves replaying existing data, then the OCP \eqref{eq:V_NLP} and possibly \eqref{eq:Q_NLP}
must be solved numerous times. The computational time spent on training the MPC scheme depends directly on the
computational time required to solve it.


To build and run the \ac{NLP} solvers, we use the \Python\, interface of \acados~\cite{verschueren2022acados}. It is in
general necessary to build two \ac{NLP} solvers, one for the \ac{NLP}~\eqref{eq:V_NLP} and one for the
\ac{NLP}~\eqref{eq:Q_NLP}. Carrying two \ac{NLP} solvers is due to the additional input constraint. However, the solution
of the \ac{NLP}~\eqref{eq:V_NLP} can be used as a good initial guess for the \ac{NLP}~\eqref{eq:Q_NLP}.

The \rlmpc\, package extracts the necessary information from the \ac{OCP} solutions from \acados\, and \CasADi\, and
provides the necessary interfaces to use the \ac{NLP} solvers in the context of training \ac{RL} algorithms. The user
defines their \ac{OCP} via an \texttt{AcadosOcp} object, which is assumed to be  well-conditioned (e.g. satisfies LICQ
and SOSC). Additional solver options can also be set. One way to generate the sensitivities is to extract the
primal-dual solution from the \texttt{AcadosOcpSolver} object, re-build the \ac{NLP} and use linear-algebra routines to
compute the sensitivities. A more tightly integrated approach is to use additional autogenerated code that allows
\acados\, to generate the sensitivities with respect to the parameters directly, which is significantly more efficient.
We discuss this next.

\subsection{Parametric Sensitivities from \acados}
The \acados\, software package provides efficient SQP-type solvers for the numerical solution of \ac{OCP} structured NLPs.
In order to efficiently compute the sensitivities of an \ac{NLP} solution and the optimal cost with respect to the parameters, we extended the \acados\, software package to provide the necessary functionality.
For the value function derivative, the Lagrange gradient with respect to the parameters $\theta$ needs to be computed.

For the policy gradient, the computation~\eqref{eq:policy_gradient} can be divided into the following steps:
\begin{enumerate}
   \item Compute the solution $y^\star$ of the NLP: This can be done with an inexact Hessian, such as a regularized Hessian, or a Gauss-Newton Hessian approximation
   \item Compute and factorize exact Hessian: The IFT requires an exact Hessian. As an efficient factorization, we use the Riccati based algorithm from \HPIPM~\cite{Frison2020a}.
   \item Evaluate the right hand side $\nabla_{\theta } \xi_\theta (s,y^\star) $, i.e. the Jacobian of the Lagrange gradient with respect to the parameters $\theta$
   \item Perform a prefactorized Riccati based solution for each parameter $\theta_i$.
\end{enumerate}
In cases where the original \ac{OCP} is convex, it can be solved with the exact Hessian in the first step, which renders
the second step unnecessary. All steps have been efficiently implemented within the \acados\, software package and can
be conveniently executed through \Python.

\subsection{Environment}

Considering the implementation and testing of \ac{RL} algorithms, there are several open libraries to use. Popular libraries include
\texttt{Gym}~\cite{brockman2016openai} and its continued maintenance in \Gymnasium~\cite{towers2023gymnasium} for building environments and
interfacing with \ac{RL} algorithms. We decided to use \Gymnasium\, as it appears to be the most actively maintained library and
de-facto standard for \ac{RL} environments. It is straightforward to implement custom environments and to interface with
\ac{RL} algorithms. The \rlmpc\, package is rather independent of the environment considering that we distinguish between the
model of the environment (i.e. the dynamics in \eqref{eq:V_constraint1}) and the environment itself (i.e. the dynamics in
\eqref{eq:dyn_mdp} or \eqref{eq:dyn_classic}). At the time of writing, we do not support vector environments.

\subsection{Reinforcement Learning Algorithms}

Similar to \ac{MPC}, for \ac{RL} algorithms, there is a range of libraries, including
\texttt{agents}\footnote{\url{https://www.tensorflow.org/agents}},
\texttt{RLlib}\footnote{\url{https://docs.ray.io/en/latest/rllib/index.html}}~\cite{liang18b}, and
\stablebaselines~\cite{raffin2021stablebaselines3}.  As a first step, we decided to use \stablebaselines\, due to its wide
range of implementations of \ac{RL} algorithms and its continued maintenance. Moreover, \stablebaselines\, is built on top of
\PyTorch~\cite{paszke2019pytorch}, which is often used in custom implementations of \ac{RL} algorithms. It
targets \ac{RL} based on \acp{NN} and provides implementations of \ac{RL} algorithms such as \ac{DPG} by
\cite{silver2014deterministic} and \ac{TD3} by \cite{fujimoto2018addressing} with \acp{NN} or \acp{MLP} as function
approximators. The \rlmpc\, package provides the necessary interfaces to use these algorithms in the context of training
with \ac{MPC} as the actor (and possibly critic). A necessary compromise is that the \rlmpc\, package needs to
"disguise" the \ac{MPC} scheme~\eqref{eq:V_NLP} as an \ac{MLP} or \ac{NN}. Instead of performing conventional forward passes,
the \ac{MPC} scheme is solved using \ac{NLP} solvers. The gradients of the \ac{NLP} solutions~\eqref{eq:V_Q_gradient},
\eqref{eq:policy_gradient} are then used to update the parameters of the \ac{MPC} scheme, whereas in conventional
\ac{DRL} algorithms, the parameters of the \ac{MLP} would be updated using backpropagation. Less modular
implementations of \ac{RL} algorithms as provided by \CleanRL~\cite{huang2022cleanrl} may be better suited, as they
allow for more direct integration of the \ac{MPC} scheme at the expense of duplicated code.

\section{Case studies}
\label{sec:case_studies}

In this section, we present two case studies: a sensitivity evaluation of the policy gradient and a Q-Learning numerical
example. The first case study in \secref{sec:sensitivity_evaluation} evaluates the computational times for computing the
policy gradient with varying number of parameters $n_\theta$ of the \ac{MPC} scheme, contrasting our extension to
\texttt{acados}\, to the use of standard linear-algebra routines in terms of computational efficiency. The second case
study in \secref{sec:q_learning_example} demonstrates the use of the \rlmpc\, package to implement a Q-Learning
algorithm for a linear system with a quadratic cost function.

\subsection{Sensitivity Evaluation of the Policy Gradient}
\label{sec:sensitivity_evaluation}

We consider a modified version of the chain-mass system in \cite{wirsching2006Fast} with $n_\mathrm{mass}$ masses
connected by massless springs and dampers. The state of the system is given by the positions $x_i$ and velocities
$\dot{x}_i$ of the $n_\mathrm{mass}-1$ masses, with fixed position of the first mass. The forces acting through the
links are given by non-isotropic springs and dampers with stiffness $k_i\in\mathbb{R}^3$, rest length
$l_i\in\mathbb{R}^3$, and damping $d_i\in\mathbb{R}^3$, with disturbances $w_i\in\mathbb{R}^3$ acting on the
$n_\mathrm{mass} - 2$ freely moving masses. The cost function is quadratic in the state and control inputs, with dense
weight matrices $Q_\theta\in\mathbb{R}^{n_x\times n_x}$ and $R\in\mathbb{R}_\theta^{n_u\times n_u}$, with $n_x,\,n_u$
denoting the respective state and input dimensions. We use this example to evaluate the computation times for the
policy gradient with varying number of parameters $n_\theta$ which is a function of $n_\mathrm{mass}$. More details can
be found in \cite{verschueren2022acados, wirsching2006Fast}.

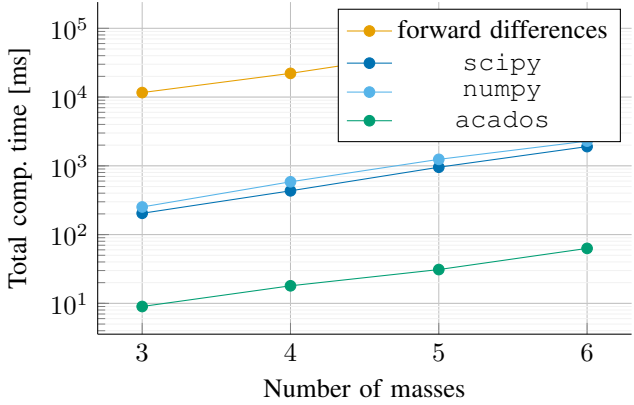
\begin{figure}[htbp]
   \centering
   \begin{tikzpicture}
      \definecolor{color1}{RGB}{0, 114, 178}
      \definecolor{color2}{RGB}{230, 159, 0}
      \definecolor{color3}{RGB}{86, 180, 233}
      \definecolor{color4}{RGB}{0, 158, 115}
      \newcommand{\height}{6cm}
      \newcommand{\linewidthinpt}{1}
      \begin{axis}
         [ymode=log, axis y line*=left, axis x line*=bottom, xtick={3,4,5,6}, xlabel={Number of masses}, ylabel={Total
                  comp. time [ms]}, height=\height, width=\columnwidth,     grid=both, grid style={line width=.1pt,
                  draw=gray!10}, major grid style={line width=.2pt,draw=gray!50},]
         \addplot[mark=*,color=color2] table [x=n, y=finitedifferences, col sep=comma] {fig/chain_mass_timings_total.csv};
         \addplot[mark=*,color=color1] table [x=n, y=numpy, col sep=comma] {fig/chain_mass_timings_total.csv};
         \addplot[mark=*,color=color3] table [x=n, y=splinalg, col sep=comma] {fig/chain_mass_timings_total.csv};
         \addplot[mark=*,color=color4] table [x=n, y=acados, col sep=comma] {fig/chain_mass_timings_total.csv};
         \legend{forward differences, \scipy, \numpy, \acados}
      \end{axis}
   \end{tikzpicture}
   \caption{Total time needed to compute the policy gradient for the chain-mass system with varying number of masses.}
   \label{fig:timings}
\end{figure}

We provide a comparison of the computation times to compute the policy gradient using different approaches.
The baseline is finite differences, which requires $n_\theta$ additional \ac{OCP} solutions to compute the policy
gradient. Then we compare re-building the NLP and utilizing general-purpose linear algebra solvers with the computation
via \acados\,, which provides a more efficient alternative. The NLP object uses \numpy\, to interface linear algebra
routines, but other libraries such as \scipy\, yield similar results as shown in \figref{fig:timings}. The computation
times are averaged over 100 OCP solutions for incrementally increasing the mass of the third ball of the chain-mass system.

The results show that using \acados\, results in significantly lower computation times compared to the other methods. This
is mainly attributed to the efficient Riccati recursion used in \acados, in addition to the compiled code without memory
allocation and panel-major linear algebra based on \texttt{BLASFEO}~\cite{frison2018blasfeo}. This makes it computationally more efficient than
the general-purpose linear algebra solvers in the libraries \numpy\, or \scipy. However, note that the \numpy/\scipy\,
implementations use the full Jacobians in \eqref{eq:KKT}. A cheaper alternative would be to use Jacobian-times-vector
products.

\subsection{Q-Learning Numerical Example}
\label{sec:q_learning_example}

We present a variant of the linear MPC example from~\cite{gros2020datadriven} to demonstrate the use of the \rlmpc\,
package. The example is based on a discrete linear time-invariant system that is subject to state and input constraints.
The \ac{OCP} is defined as
\begin{subequations}
   \begin{align}
      \min _{x,u} & \,\, V_0 + \frac{\gamma ^N}{2}x_N^\top S_N x_N + \sum _{k=0}^{N-1}f^\top\left[\begin{array}{c}x_k \\ u_k\end{array}\right]\nonumber                                      \\
                  & + \sum _{k=0}^{N-1} \frac{1}{2}\gamma ^k\left(\left\Vert x_k\right\Vert ^2 + \left\Vert u_k\right\Vert ^2 + w^\top s_k\right)  \label{eq:lti_stage_cost}                 \\
      \text{s.t.} & \quad x_{k+1} = Ax_k + Bu_k + b                                                                                                                                          \\
                  & \quad \left[\begin{array}{c}\phantom{-}0 \\ -1\end{array}\right] -s_k\leq x_k \leq \left[\begin{array}{c}1 \\ 1\end{array}\right] + s_k \label{eq:lti_state_constraints} \\
                  & \quad -1\leq u_k \leq 1, \qquad x_0 = s,
   \end{align}
\end{subequations}
and the following parameters can be modified by the RL algorithm:
\begin{align}
   \theta = \left(V_0,\,b,\,f,\, A,\,B\right)
\end{align}
The MPC scheme uses a horizon of $N=40$ and a discount factor of $\gamma=0.9$. The terminal cost weights $S_N$ are set
to the solution of the discrete-time algebraic Riccati equation. The weights for the constraint relaxation $w$ are
set to $w^\top=[100, 100]$. Minimizing \eqref{eq:lti_stage_cost} drives the system to the nominal steady state at
the origin where the lower constraints of the first state are active~\eqref{eq:lti_state_constraints}.

The initial model of the system is given by the matrices
\begin{align}
   A = \left[\begin{array}{cc}1 & 0.25\\ 0 & 1 \end{array}\right],\quad B= \left[\begin{array}{c}0.0312\\ 0.25\end{array}\right]
\end{align}
and all other parameters are initially set to zero.

The true system is following the dynamics
\begin{align}
   x_{k+1} = \left[\begin{array}{cc}0.9 & 0.35\\ 0 & 1.1 \end{array}\right]x_k + \left[\begin{array}{c}0.0813\\ 0.2\end{array}\right]u_k + \left[\begin{array}{c}e_k \\ 0\end{array}\right],
\end{align}
where $e_k$ is a random, uncorrelated disturbance sampled from a uniform distribution on $[-0.1,0]$. It will therefore
drive the first state to the lower bound and lead to state-constraint violations.

\begin{figure}[htbp]
   \centering
   \begin{adjustbox}{max width=\columnwidth}
      \begin{tikzpicture}

    \definecolor{color1}{RGB}{0, 114, 178}
    \definecolor{color2}{RGB}{230, 159, 0}
    \definecolor{color3}{RGB}{0, 158, 115}
    \definecolor{color4}{RGB}{86, 180, 233}
    \newcommand{\height}{3.5cm}
    \newcommand{\linewidthinpt}{1}
    \begin{groupplot}[group style={group size=1 by 5, vertical sep=0.5cm}, height=\height, width=\columnwidth, xmin=0, xmax=30, grid=both, axis y line*=left, axis x line*=bottom]
        \nextgroupplot[
            xlabel={},
            ylabel={$B_i$},
            height=\height,
            width=\columnwidth,
            axis y line*=left,
            axis x line*=bottom,
            ymin=-0.05,
            ymax=+0.26,
            line width = \linewidthinpt pt,
            xticklabels={},
        ]
        \addplot[color1, no markers, solid] table[x=episode, y=B_0,col sep=comma] {fig/linear_system_mpc_qlearning/data.csv};
        \addplot[color2, no markers, solid] table[x=episode, y=B_1,col sep=comma] {fig/linear_system_mpc_qlearning/data.csv};
        \nextgroupplot[
            xlabel={},
            ylabel={$b_i$},
            height=\height,
            width=\columnwidth,
            axis y line*=left,
            axis x line*=bottom,
            ymin=-0.1,
            ymax=+0.01,
            ytick={0,-0.05, -0.1},
            yticklabels={0, -0.05, -0.1},
            line width = \linewidthinpt pt,
            xticklabels={},
        ]
        \addplot[color1, no markers, solid] table[x=episode, y=b_0,col sep=comma] {fig/linear_system_mpc_qlearning/data.csv};
        \addlegendentry{$b_1$}
        \addplot[color2, no markers, solid] table[x=episode, y=b_1,col sep=comma] {fig/linear_system_mpc_qlearning/data.csv};
        \addlegendentry{$b_2$}
        \nextgroupplot[
            xlabel={},
            ylabel={$f_i$},
            height=\height,
            width=\columnwidth,
            axis y line*=left,
            axis x line*=bottom,
            ymin=-0.001,
            ymax=+0.003,
            line width = \linewidthinpt pt,
            xticklabels={},
        ]
        \addplot[color1, no markers, solid] table[x=episode, y=f_0,col sep=comma] {fig/linear_system_mpc_qlearning/data.csv};
        \addlegendentry{$f_1$}
        \addplot[color2, no markers, solid] table[x=episode, y=f_1,col sep=comma] {fig/linear_system_mpc_qlearning/data.csv};
        \addlegendentry{$f_2$}
        \addplot[color3, no markers, solid] table[x=episode, y=f_2,col sep=comma] {fig/linear_system_mpc_qlearning/data.csv};
        \addlegendentry{$f_3$}
        \nextgroupplot[
            xlabel={},
            ylabel={$V_0$},
            height=\height,
            width=\columnwidth,
            axis y line*=left,
            axis x line*=bottom,
            line width = \linewidthinpt pt,
            xticklabels={},
        ]
        \addplot[color1, no markers, solid] table[x=episode, y=V_0,col sep=comma] {fig/linear_system_mpc_qlearning/data.csv};
        \nextgroupplot[
            xlabel={Episode},
            ylabel={$J(\pi^\mathrm{MPC})$},
            line width = \linewidthinpt pt,
            ymode=log
        ]
        \addplot[color1, no markers, solid] table[x=episode, y=cost,col sep=comma] {fig/linear_system_mpc_qlearning/data.csv};
    \end{groupplot}
\end{tikzpicture}
   \end{adjustbox}
   \caption{Evolution of the parameters and the accumulated cost for each episode under the MPC-based policy $\pi^\mathrm{MPC}$.}
   \label{fig:training}
\end{figure}
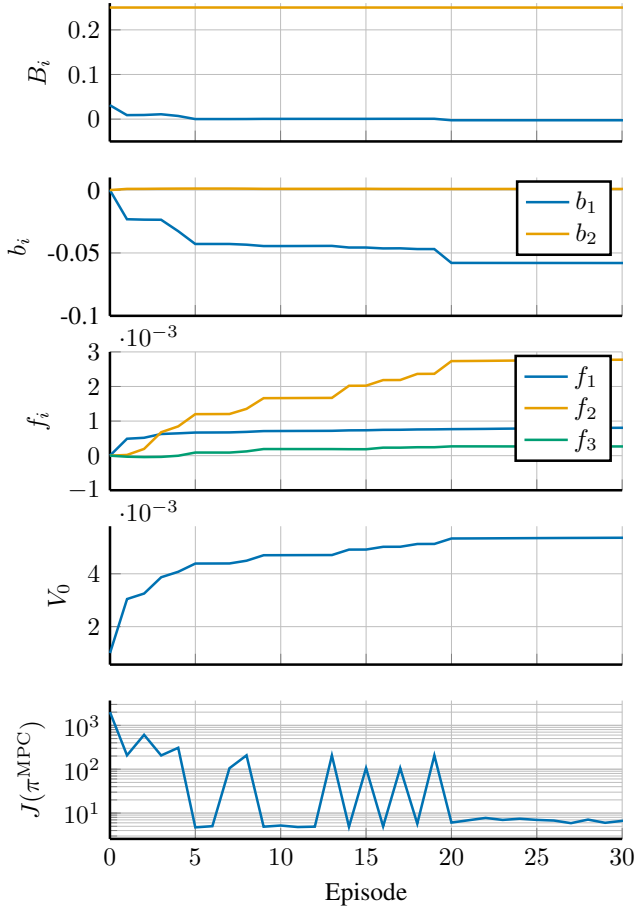

We use the \rlmpc\, package to train the \ac{MPC} using Q-Learning. The training phase is carried out for 30 episodes, each
consisting of 100 time steps. The learning rate is set to $\alpha=10^{-4}$ and we do not use any exploration strategy.
The initial and final episodes are shown in \figref{fig:episodes_before_after}. The training progress is shown in
\figref{fig:training}, where we only report the results until the 30th episode given that the learning converges after
20 episodes. As in the original example, the \ac{MPC} scheme is trained to avoid state-constraint violations by backing
off from the nominal steady state at the origin. The RL algorithm does this mainly by adapting the cost parameters $V_0$
and $f$, and the bias term $b$ in the model. The state trajectories are quickly driven to the reference and the
accumulated stage cost in each episode is dominated by penalties for state-constraint violations. An interesting
observation from \figref{fig:training} is that constraint violations are avoided as soon as the bias parameter $b$
converges to a value that slightly exceeds the expected value of the disturbance $e_k$ in the true system. Also note
that one would normally use longer episode lengths and a smaller learning rate for a more stable learning process with a
monotonic decrease in the accumulated cost.

\begin{figure}[htbp]
   \centering
   \begin{adjustbox}{max width=\columnwidth}
      \begin{tikzpicture}

    \definecolor{color1}{RGB}{0, 114, 178}
    \definecolor{color2}{RGB}{230, 159, 0}
    \definecolor{color3}{RGB}{86, 180, 233}
    \definecolor{color4}{RGB}{0, 158, 115}
    \newcommand{\height}{3cm}
    \newcommand{\linewidthinpt}{1}
    \begin{groupplot}[group style={group size=1 by 4, vertical sep=0.25cm}, height=\height, width=\columnwidth, xmin=0, xmax=100, grid=both, axis y line*=left, axis x line*=bottom]
        \nextgroupplot[
            xlabel={},
            ylabel={$s_1$},
            height=\height,
            width=\columnwidth,
            axis y line*=left,
            axis x line*=bottom,
            ymin=-0.1,
            ymax=+0.8,
            line width=\linewidthinpt pt,
            xticklabels={},
        ]
        \addplot[color1, no markers, solid] table[x=k, y=x_0,col sep=comma] {fig/linear_system_mpc_qlearning/episode_0.csv};
        \addplot[color2, no markers, solid] table[x=k, y=x_0,col sep=comma] {fig/linear_system_mpc_qlearning/episode_30.csv};
        \legend{Episode 1, Episode 30}
        \addplot[black, no markers, dashed] coordinates {(0,0) (100,0)};
        \nextgroupplot[
            xlabel={},
            ylabel={$s_2$},
            height=\height,
            width=\columnwidth,
            ymin=-0.2,
            ymax=+0.8,
            line width=\linewidthinpt pt,
            xticklabels={},
        ]
        \addplot[color1, no markers, solid] table[x=k, y=x_1,col sep=comma] {fig/linear_system_mpc_qlearning/episode_0.csv};
        \addplot[color2, no markers, solid] table[x=k, y=x_1,col sep=comma] {fig/linear_system_mpc_qlearning/episode_30.csv};
        \nextgroupplot[
            xlabel={},
            ylabel={$u$},
            ymin=-1,
            ymax=+1,
            xticklabels={},
            line width=\linewidthinpt pt
        ]
        \addplot[color1, no markers, solid] table[x=k, y=u,col sep=comma] {fig/linear_system_mpc_qlearning/episode_0.csv};
        \addplot[color2, no markers, solid] table[x=k, y=u,col sep=comma] {fig/linear_system_mpc_qlearning/episode_30.csv};
        \nextgroupplot[
            xlabel={$k$},
            ylabel={$L$},
            ymax=+200,
            ymode=log,
            line width=\linewidthinpt pt
        ]
        \addplot[color1, no markers, solid] table[x=k, y=cost,col sep=comma] {fig/linear_system_mpc_qlearning/episode_0.csv};
        \addplot[color2, no markers, solid] table[x=k, y=cost,col sep=comma] {fig/linear_system_mpc_qlearning/episode_30.csv};

    \end{groupplot}
\end{tikzpicture}
   \end{adjustbox}
   \caption{Results before and after training. State-constraint violations and corresponding
      large control actions are avoided after training by backing off the constraint (and reference) at the origin.}
   \label{fig:episodes_before_after}
\end{figure}
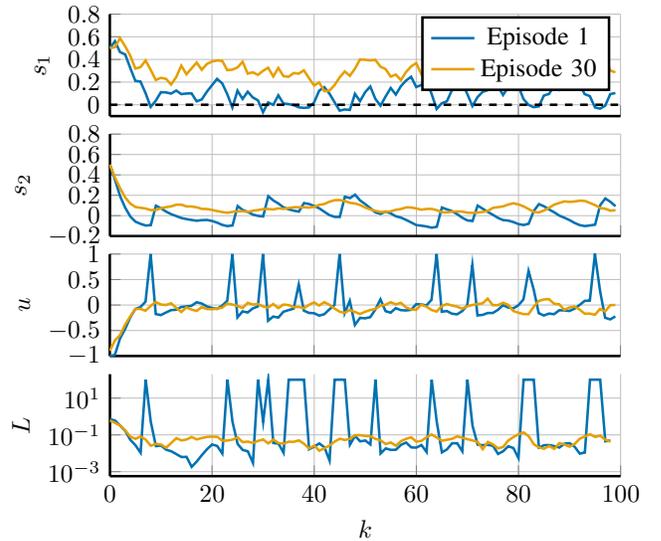

\section{Future Work}
\label{sec:future_work}

In this section, we discuss future work and potential extensions of the \rlmpc\, package.

\subsection{Reinforcement Learning Algorithms}

The \rlmpc\, package is built around \stablebaselines, which is built on top of \PyTorch. All \ac{RL} algorithms in \stablebaselines\, are based on \acp{NN} or \acp{MLP}, and some of them are limited to discrete action spaces, e.g. \ac{DQN}.
For simple algorithms such as Q-Learning, it is straightforward to implement custom implementations instead, as we have
done in the example in \secref{sec:case_studies}. However, for more complex algorithms such as \ac{DPG} or \ac{TD3}, it
is beneficial to connect the \ac{MPC} scheme to the \ac{RL} algorithm in a more direct way.

\subsection{Parallel Sensitivity Evaluation}

The results in \secref{sec:case_studies} show that the computation of the policy gradients can be significantly
accelerated by computing the sensitivities using \acados\, directly.  Considering that learning algorithms typically
operate on a batch of samples stored in the replay buffer, an efficient implementation furthermore requires parallel
sensitivity computation for varying parameters. The resulting speedup would be beneficial for the training phase, both
for \ac{RL} algorithms and other methods requiring \ac{NLP} solution sensitivities.

\subsection{Warm Starting}

The \rlmpc\, package currently only stores the minimum information for \ac{RL} algorithms in the replay buffer, i.e. the
state, action, cost, and next state. Extending the replay buffer to store the primal-dual solution of the \ac{NLP} would
enable warm starting during the training phase and result in a more efficient training process. Considering the
case of multiple solvers running in parallel, coordinating the sampling in a way that each solver can stay close to the
previous solution for warm-starting would also be beneficial.

\subsection{Compare Computational Efficiency}

The project \diffmpc\,~\cite{amos2018differentiable} or a subset of solvers available in \CasADi\, as described
in~\cite{andersson2018sensitivity}, take a similar approach, but re-use matrix factorizations from the forward pass to
compute the sensitivities. A comparison of the computational efficiency of \acados\, and \diffmpc\, is the subject of
ongoing work.


\subsection{Additional Examples}

An extensive body of examples has appeared in the literature that use \ac{RL} based on \ac{MPC} algorithms, with
applications given in~\cite{cai2021mpcbased,baharikordabad2021mpcbased}, but also more theoretical developments
in~\cite{zanon2021safe}. The implementations in these papers are often based on custom implementations of \ac{RL}
algorithms and are not publicly accessible. We aim to reproduce these examples using the \rlmpc\, package and to provide
additional examples that demonstrate the use of the package in a wide range of control problems and for different
\ac{MPC} formulations.

\section{Conclusion}
We introduced the \rlmpc\, package, a framework designed to seamlessly integrate
\ac{MPC} schemes with \ac{RL} algorithms. This package leverages the advanced capabilities of the \acados\, OCP solvers
to connect to available \ac{RL} libraries, thereby facilitating the application of \ac{MPC} in the reinforcement learning
context. Through a straightforward example, we showcased the package's utility by training an \ac{MPC} scheme via
Q-Learning.

The \rlmpc\, package represents an initial step to open-source implementations of \ac{MPC}-based \ac{RL} algorithms. We
showed that computing the solution sensitivities of an OCP can be performed more than a magnitude faster with tailored
solvers compared to general purpose libraries. Our future objectives target the expansion of this package to
accommodate a broader spectrum of RL algorithms. Moreover, we aim to provide additional examples that demonstrate the
use of the package in a broad range of control problems and for different \ac{MPC} formulations, and provide a comparison
of the computational efficiency with alternative packages such as \diffmpc.
\label{sec:conclusion}

\newpage

\bibliography{references}

\end{document}